\documentclass[12pt]{article}
\usepackage[dvips]{graphicx}

\usepackage{epsfig,amsmath,amssymb,verbatim,mathrsfs}

\usepackage{slashed}
\usepackage{graphicx}
\usepackage{amsmath}
\usepackage{amssymb}
\usepackage{epstopdf}
\usepackage{verbatim}
\usepackage{url}
\usepackage{amsfonts}
\usepackage{layout}
\usepackage{slashed}
\usepackage[usenames,dvipsnames]{color}

\usepackage{hyperref}	

\newcommand{\drawsquare}[2]{\hbox{%
\rule{#2pt}{#1pt}\hskip-#2pt
\rule{#1pt}{#2pt}\hskip-#1pt
\rule[#1pt]{#1pt}{#2pt}}\rule[#1pt]{#2pt}{#2pt}\hskip-#2pt
\rule{#2pt}{#1pt}}

\newcommand{\fund}{~\raisebox{-.5pt}{\drawsquare{6.5}{0.4}}~}
\newcommand{\antifund}{~\overline{\raisebox{-.5pt}{\drawsquare{6.5}{0.4}}}~}
\newcommand{\antisym}{~\raisebox{-3.5pt}{\drawsquare{6.5}{0.4}}\hskip-6.9pt%
        \raisebox{3pt}{\drawsquare{6.5}{0.4}}~}

\newcommand{\beq}{\begin{equation}}
\newcommand{\eeq}{\end{equation}}
\newcommand{\bea}{\begin{eqnarray}}
\newcommand{\eea}{\end{eqnarray}}

\newcommand{\mr}[1]{\mathrm{#1}}

\newcommand{\OO}{\mathcal{O}}

\newcommand{\be}{\begin{equation}}
\newcommand{\ee}{\end{equation}}

\begin{document}

\setlength{\baselineskip}{0.2in}


\begin{titlepage}
\noindent
\begin{flushright}
SLAC-PUB-14930
\end{flushright}
\vspace{1cm}

\begin{center}
  \begin{Large}
    \begin{bf}

A test for emergent dynamics

     \end{bf}
  \end{Large}
\end{center}
%

\begin{center}

{\bf Anson Hook}\\
\vspace{.5cm}
  \begin{it}
Theory Group, SLAC National Accelerator Laboratory, \\
2575 Sand Hill Rd, Menlo Park, CA 94025\\
\vspace{0.5cm}
\end{it}

\end{center}


\begin{abstract}

A generalization of $a$-maximization is proposed that maximizes $a$ subject to inequalities rather than equalities.  An implication of this conjecture is that in the absence of emergent symmetries, there is a maximum R-charge for fields appearing in the path integral.  This maximum R-charge leads to a novel way of detecting emergent Abelian symmetries and non-Abelian gauge symmetries.

 \end{abstract}

\vspace{1cm}

\end{titlepage}

\setcounter{page}{2} 


\section{Introduction\label{sec:introduction}}

Supersymmetric gauge theories have many possible infrared (IR) phases.  Given an ultraviolet (UV) theory, it is not known in general what the IR phase is.  More can be said about the phase structure if there is a dual description\cite{Seiberg:1994pq}.  Assuming that there are no emergent symmetries and that the IR is a conformal field theory (CFT), $a$-maximization can be used to determine its properties\cite{Intriligator:2003jj}.  Flow arguments can then be used to argue what the IR theory is\cite{Barnes:2004jj,Intriligator:2003mi}.

$a$-maximization is therefore an appealing tool by which to study the IR phase of supersymmetric theories.  
By studying the R-charges of the chiral ring operators, one can impose constraints such as unitarity to check if the proposed IR theory makes sense.  If the R-charge of a gauge invariant operator is less than $2/3$, there must be an emergent symmetry which enforces unitarity.

$a$-maximization can also be done using Lagrange multipliers to enforce anomaly free conditions and marginality of superpotential terms.  Rather than being a simple trick used to find critical points, \cite{Kutasov:2003ux,Kutasov:2004xu} motivated giving the Lagrange multipliers a physical interpretation.  Near a free fixed point, it was shown that Lagrange multipliers are proportional to the coupling constants squared.  R-charges as a function of Lagrange multipliers were interpreted as RG flow.


In this paper, a generalization of $a$ maximization is proposed where $a$ is maximized subject to inequalities rather than equalities.  The inequalities imposed are that the superpotential terms are marginal or irrelevant and that gauge groups are IR free or conformal. Many interesting properties emerge from this conjectured generalization of $a$-maximization.

When maximizing a function subject to equalities, Lagrange multipliers can be used and obey the Lagrange conditions.  When maximizing a function subject to inequalities new conditions for the Lagrange multipliers exist, the Karush-Kuhn-Tucker conditions\cite{KKT}.  One of these conditions is that the Lagrange multipliers are positive.

As a direct result of the positivity of Lagrange multipliers, it is shown that barring emergent symmetries the R-charge for gauge singlet ``fundamental" fields\footnote{The notation of ``fundamental" will refer to fields which are integrated over in the path integral and not any notion of being a free field in the UV.  These ``fundamental" fields are those which appear in the $a$-maximization procedure.}, $\Phi$, appearing in the path integral obeys the constraints $2/3 \le R_\Phi \le 4/3$.  For other recent work on emergent symmetries see \cite{Buican:2011ty}.

In Sec.~\ref{sec:amax}, $a$-maximization and Lagrange multipliers are reviewed.  In Sec.~\ref{sec:improve}, an extension to $a$-maximization is proposed.  In Sec.~\ref{sec:emergent}, a maximum R-charge for gauge singlet ``fundamental" fields is proven, which allows for several new ways to detect emergent symmetries.  Finally in Sec.~\ref{sec:tests}, some basic tests of the proposal are applied.  The appendix describes a specific application of these results to the deconfinement trick.

\section{$a$-maximization}\label{sec:amax}

This section gives a brief review of $a$-maximization and identifies a trend whereby negative Lagrange multipliers imply unphysical results.  Assume that we are given a theory where it is known which superpotential terms and gauge groups are part of the IR CFT.  The interacting superpotential is $W = \sum_a \prod_i \Phi_i^{n_{a i}}$ for fields $\Phi_i$ with R-charge $R_i$ and the non-IR free gauge groups are $G_g$.  $a$-maximization is the statement that the correct R-symmetry is the symmetry which maximizes the function
\bea
a_\mr{trial} &=& \sum_g 2 |G_g| + \sum_i \mr{dim}(i) (3 (R_i-1)^3 - (R_i-1))
\eea
subject to the constraints
\bea
  \sum_i n_{a i} R_i = 2 &\qquad& \forall a \nonumber \\
  T_{G_g} + \sum_i T_{g,r_i} (R_i-1) = 0 &\qquad& \forall g
\eea
These constraints can be implemented using Lagrange multipliers.  In what follows we do not adopt the interpretation advocated by \cite{Kutasov:2003ux};  rather we will use the original version of $a$-maximization but find solutions using Lagrange's method.  We identify the trend that {\it negative Lagrange multipliers are found to result from unphysical restrictions}.  The convention used for the sign of the Lagrange multipliers is that $a$ is modified to
\bea
a = a_\mr{trial} 
  - \sum_a \lambda_a (2 - \sum_i n_{a i} R_i)  - \sum_g \lambda_g (T_{G_g} + \sum_i T_{g,r_i} (R_i-1)) 
\eea

To obtain an idea of what a positive or negative Lagrange multiplier entails, consider taking a CFT and adding a new interaction  $W = \prod_i \Phi_i^{m_i}$.  
\bea
a_\mr{new} = a_\mr{old} - \lambda (2 - \sum_i m_i R_i)
\eea
we can force this superpotential term to be part of the IR CFT and look for a new fixed point.  If this interaction shifts the location of the maximum only by a small amount $R_i = R^\mr{CFT}_i + \epsilon_i$, one can Taylor expand for the leading order corrections.  Using the criticality condition, one obtains
\bea
\sum_i \epsilon_i \frac{d a_\mr{new}}{d R_i} &=& \sum_i \epsilon_i (\frac{d a_\mr{old}}{d R_i} \big |_{R^\mr{CFT} + \epsilon} + \lambda m_i)  \nonumber \\
&=& \sum_{i,j} \epsilon_i \frac{d^2 a_\mr{old}}{d R_i d R_j} \big |_{R^{CFT}} \epsilon_j + \lambda \sum_i \epsilon_i m_i + \OO(\epsilon^3) \\
&=& 0 \nonumber \\
\lambda \sum_i \epsilon_i m_i &=& - \sum_{i,j} \epsilon_i \frac{d^2 a_\mr{old}}{d R_i d R_j} \big |_{R^{CFT}} \epsilon_j > 0
\eea
where the last step follows from the negative definiteness of $\frac{d^2 a_\mr{old}}{d R_i d R_j} \big |_{R^{CFT}}$; a result that follows from $R^{CFT}$ maximizing the original $a$.  From this, we get $\text{sign}(\lambda) = \text{sign}(\sum_i m_i \epsilon_i)$.  Marginality of the superpotential gives $\sum_i m_i \epsilon_i = 2 - \sum n_i R_i^\mr{CFT}$.  We see that $\lambda$ is negative for irrelevant interactions and positive for relevant interactions.  In this example, negative Lagrange multipliers emerge when trying to undertake the unphysical step of making an irrelevant operator marginal\footnote{Negative Lagrange multipliers can also be obtained when making a relevant coupling marginal.  An example of this is magnetic SQCD in the free electric window.  More details are given in Sec.~\ref{sec:method1}}.

The pathologies associated with negative Lagrange multipliers can be made even more explicit by considering several more examples.  Consider SQCD with $N_f > 3 N_c$;  one can maximize $a$ while forcing the beta function of the gauge group to vanish.  The result is a negative Lagrange multiplier with quarks having $R_Q > 2/3$.  It is known that there is no unitary interacting conformal fixed point and that there is instead a free fixed point.  In this example, a negative Lagrange multiplier indicated that the conformal theory found by $a$-maximization was not physical.

Another set of examples are the theories from \cite{Intriligator:2003mi}.  They examined $SU(N_c)$ with $N_f$ flavors, two adjoints and various superpotential terms.  The theories were classified as $A$,$E$,$D$ and $O$ type theories based on their superpotentials.  By analyzing the flow, they determined that for certain ranges of $x = N_c/N_f$, it was impossible to flow to some of these theories despite the fact that $a$-maximization still yielded a consistent fixed point.  For example, they found that their $E_7$ theory could no longer be reached by RG flow for $x > \sqrt{17}$ in the large $N$ limit.  As such, they claimed that the fixed point ceased to exist at this point.  By computing the Lagrange multiplier, we see that at this exact point is when the Lagrange multiplier goes negative.  For the rest of their examples, it is simple to check that when the theories cease to exist, a Lagrange multiplier is going negative.

Near a free fixed point, \cite{Kutasov:2003ux} showed that Lagrange multipliers $\lambda \sim \alpha + \OO(\alpha^2)$ and are therefore always positive.  Negative Lagrange multipliers would require imaginary gauge couplings.  The $a$ theorem for theories near a free fixed point also requires that Lagrange multipliers flow from zero to a positive value\cite{Barnes:2004jj}.  These arguments for positive Lagrange multipliers are perturbative in nature.  The high order corrections to $\lambda \sim \alpha  + \OO(\alpha^2)$ could be of any sign and could dominate when the gauge coupling becomes large.

While this section did not prove that Lagrange multipliers must be positive, it is plausible that they should always be positive in physical scenarios.  Their positivity motivates a simple extension to $a$-maximization put forth in Sec.~\ref{sec:improve}.

\section{$a$-maximization subject to inequalities}\label{sec:improve}

The correct R-charges are obtained by maximizing $a$ subject to the constraints $R(W)=2$ and $\beta_g = 0$.  These constraints are only imposed on the marginal superpotential terms and on the non-IR free gauge groups.

Consider maximizing $a$ instead subject to the {\it inequalities} $2-R(W) \le 0$ and $T_{G_g} + \sum_i T_g(i) (R_i-1) \le 0$,  i.e. maximize $a$ subject to the constraint that all the superpotential terms are either irrelevant or marginal and that all gauge groups are IR free or conformal.  The inequality does not require one to assume which superpotential terms are marginal and which are irrelevant;  it is taken care of by the maximization procedure.

Maximizing a function subject to inequalities gives the following Karush-Kuhn-Tucker conditions.  Take the function 
\bea\label{eq:atrial}
a_\mr{trial} &=& \sum_g 2 |G_g| + \sum_i \mr{dim}(i) (3 (R_i-1)^3 - (R_i-1))  \nonumber \\
  &-& \sum_a \lambda_a (2 - \sum_i n_{a i} R_i)  - \sum_g \lambda_g (T_{G_g} + \sum_i T_{g,r_i} (R_i-1)) 
\eea
A necessary condition for the existence of a maximum in the allowed region, is that there exist variables $\lambda_a , \lambda_g$ such that
\bea
\frac{d a_\mr{trial}}{d R_i} &=& 0 \label{eq:KKT1}\\
 2 - \sum_i n_{a i} R_i \le 0 &\qquad& T_{G_g} + \sum_i T_{g,r_i} (R_i-1) \le 0 \label{eq:KKT2}\\
\lambda_a , \lambda_g &\ge& 0 \label{eq:KKT3}\\
 \lambda_a (2 - \sum_i n_{a i} R_i) &=&  \lambda_g (T_{G_g} + \sum_i T_{g,r_i} (R_i-1))  = 0 \label{eq:KKT4}
\eea
Of course one must still check that the solution found is a maximum.  Interested readers can find their proof in \cite{KKT}.  Eq.~\ref{eq:KKT1} follows from looking for a critical point while Eq.~\ref{eq:KKT2} is simply the inequality to be satisfied.   Eq.~\ref{eq:KKT3} and Eq.~\ref{eq:KKT4} are the result of several pages of nonlinear programming.

Eq.~\ref{eq:KKT4} is the statement that an irrelevant superpotential corresponds to a zero Lagrange multiplier.  Thus, no constraint is enforced for irrelevant superpotentials.  Exactly marginal superpotentials also have Lagrange multipliers of zero, as shown in Sec.~\ref{sec:amax}.

This conjectured modification of $a$-maximization gives some well motivated effects.  As mentioned previously, the heuristic measurement by which people looked for fixed points was to impose $a$-maximization, Eq.~\ref{eq:KKT1}, and look for solutions where Eq.~\ref{eq:KKT2} is satisfied.   By incorporating these inequalities into the maximization procedure, the whole approach of using flows is automatically included.

An attractive feature of this maximization procedure is the requirement that all Lagrange multipliers must be positive.  Sec.~\ref{sec:amax} argued that Lagrange multipliers should be positive.  Here we see that this result directly falls out of the maximization procedure.


\subsection{Relation to a flow analysis}

A deficit of $a$-maximization is that it assumes that one knows precisely what the symmetries of the IR fixed point are.  In the case where the emergent symmetries are simply due to superpotential terms flowing to zero, a flow analysis can be used to determine what the IR fixed point is.

The spirit of a flow analysis is as follows.  Varying holomorphic superpotential parameters does not induce phase transitions except on codimension 1 surfaces.  While not inducing a phase transition is not the same as having the same low energy physics, let us assume that the low energy physics does not change as holomorphic parameters are varied.  There are no known counter-examples to this assumption.  By making superpotential or gauge couplings very small, one can approach different fixed points before being pushed away by relevant terms.  Eventually one flows to a fixed points where all of the superpotential terms are marginal or irrelevant.

Given a fixed superpotential, there can be several solutions in which different superpotential terms are marginal with all non-marginal superpotential terms irrelevant.  For example, the solution where all terms in the superpotential are assumed to be marginal is always a consistent solution because none of the superpotential terms are relevant.  By the previous assumption, only one of these CFTs are real, the rest are not consistent field theories.  It was conjectured that the CFT with the largest $a$ value is the correct CFT\cite{Intriligator:2005if}.

Generally, sifting through the flows between fixed points is a very complicated process as turning on one coupling may cause another to flow to zero.  Given N superpotential terms and M gauge groups, one must go through $\OO(2^{M+N})$ fixed points to find the correct IR fixed point.

When maximizing $a$ subject to inequalities, these flows are automatically taken into account.  While for humans, maximizing subject to inequalities results in a similarly large number of steps as the flow process, it is a 1 step process for computers and easily automated.

\section{Detecting Emergent Gauge Symmetries with a Maximum R-Charge} \label{sec:emergent}

In this section, the results from the previous section are put to use to discover emergent Abelian or non-Abelian symmetries.  There are two ways in which the results from the previous sections can be used to detect emergent symmetries:  In the first case, operators {\bf not} part of the chiral ring hit the $R=2/3$ bound.  This case is related to the maximum R-charge for fields.  The second case is more standard where operators part of the chiral ring hit the unitarity bound of $2/3$.  

Consider magnetic SQCD. This is the theory with gauge group $SU(\tilde{N}_c)$ with $N_f$ flavors, $N_f^2$ meson fields and superpotential term $W= q M \overline{q}$.   The operator $q \overline{q}$ does not need to obey the unitarity relation $R > 2/3$.  The F term equations of motion for M set $q \overline{q} =0$ while including the Kahler potential gives $q \overline{q}$ being proportional to a bunch of derivatives.  Using the state operator correspondence for CFTs, we see that the state created by $q \overline{q}$ is a descendant and does not obey $\Delta = 3/2 R$.

This realization leads to the following interesting observation about $a$-maximization and SQCD.  Unitarity says that the meson must have R-charge $\ge 2/3$.  Applying this criteria to electric SQCD, an emergent symmetry must appear when $N_f = \frac{3}{2} N_c$.  The meson is going free at this point as signaled by unitarity.  In the magnetic theory, there is no distinct signature that something is going wrong at $N_f = \frac{3}{2} \tilde{N}_c$ as unitarity does not apply to the meson $q \tilde{q}$;  it does not tell the magnetic theory that the electric theory is going free.  As will be described in Sec.~\ref{sec:method1}, $a$-maximization subject to inequalities gives the magnetic theory a way to see that the electric theory is going free.  A flow analysis can also be used to see the same result.

\subsection{A maximal R-charge}

Using the proposed generalization of $a$-maximization, it is proven that there is a maximal R-charge for gauge singlet ``fundamental" fields.  To reiterate ``fundamental" means fields which appear in the path integral and does not refer to UV physics.  It is also assumed that fields appear in the superpotential with only positive power.  Destabilizing the origin typically higgses the conformal gauge group and invalidates many of the symmetries used for $a$-maximization.

When solving for the Lagrange multipliers of a gauge singlet field, we must solve $\frac{d a}{d R_i} =0$ with $a$ given in Eq.~\ref{eq:atrial}.  Assuming a superpotential $W=\sum_a \prod_i \Phi_i^{n_{a i}}$ and  solving for the R-charges of fields yields
\bea
R_i = 1 \pm \frac{1}{3} \sqrt{1 - \sum_a \frac{n_{a i}}{|r_i|} \lambda_a }
\eea
One then sees that because the Lagrange multipliers are positive,  
\bea
\frac{2}{3} \le R_i \le \frac{4}{3}
\eea
in other words, there is a maximum R-charge for gauge singlet ``fundamental" fields.  This maximal value is an algebraic result stemming from requiring positive Lagrange multipliers and has nothing to do with any physical interpretation of $R(\lambda)$ or of $\lambda$ itself.  The unitarity constraint $R \ge 2/3$ is automatically incorporated.  The gauge Lagrange multiplier $\lambda_g$ appears in the square-root with an opposite sign for fields charged under the gauge group.

There are various uses for this new fact.  If one wishes to take a theory with no known Lagrangian description and guess a Lagrangian, then all chiral operators with R-charge greater than $4/3$ must in fact be composite operators.  They cannot be consistently made into ``fundamental" fields.  The fact that operators with R-charge greater than $4/3$ must be composite operators lets one learn about dual description of field theories.

%

\subsection{A non-solution as an indicator of emergent symmetries}\label{sec:method1}

The first and easiest method for detecting the emergence of symmetries is that there is no solution to the modified $a$-maximization procedure.  No solution should be interpreted as indicating that there is no consistent CFT that exists given the UV symmetries.  An example of this is magnetic SQCD.

Applying the results of Sec.~\ref{sec:improve}, we see that for $N_f \ge 3 \tilde{N}_c$, the Lagrange multiplier for the gauge group hits 0 and the gauge group becomes free.  For $\frac{3}{2} \tilde{N}_c < N_f < 3 \tilde{N}_c$, the Lagrange multipliers are positive indicating that a conformal fixed point is reached.  The Lagrange multiplier for the superpotential vanishes for $N_f = \frac{3}{2} \tilde{N}_c$.

For $N_f  < \frac{3}{2} \tilde{N}_c$ there is no solution.  When $N_f < \frac{3}{2} \tilde{N}_c$, the magnetic quarks have R-charge $< \frac{1}{3}$.  If the meson M is free, then the superpotential term is relevant.  If the term becomes marginal then the R-charge of M is greater than 4/3 which is impossible with positive Lagrange multipliers.  Hence there is no solution.  

A gauge singlet hitting the $4/3$ bound appears when it is coupled to a CFT via $\OO M$ and the operator $\OO$ is going below the unitarity bound. If $\OO$ had R-charge less than 2/3 then a negative Lagrange multiplier would result.  Thus we find that in the absence of emergent symmetries {\it chiral operators removed from the chiral ring by an F term still need to have $R_\OO \ge 2/3$.}

The emergent symmetry that appears when a field hits $R=4/3$ can be discovered by using flows.  Start with the theory where the superpotential $\OO M$ is not part of the CFT.  We see that there is an emergent $U(1)$ necessary to give the field $\OO$ R-charge equal to 2/3.  Once the superpotential $\OO M$ is turned on, this $U(1)$ instead rotates the fields $\OO$ and M in opposite directions.  In the case of magnetic SQCD, the emergent symmetry can be made explicit by dualizing back to electric SQCD.  The emergent symmetry is the $U(1)$ acting on the massive fields $M$ and $q \overline{q}$.  In this manner, it is possible to use the method of flows to rederive the $R=4/3$ bound.

Much like the unitarity bound, if there is no solution then an emergent $U(1)$ or more appears.  When used in conjunction with the method in Sec.~\ref{sec:method2}, it provides a very good determinant of when emergent dynamics appear.  However as shown in Sec.~\ref{sec:adjoint}, there are still emergent dynamics missed by these methods.

\subsection{Probing baryons} \label{sec:method2}

The second method for detecting emergent symmetries is more involved.  It is specifically aimed at trying to detect when emergent {\it non-Abelian gauge} symmetries go free at the same time as another field goes free.  Like all other methods, it is by no means fool-proof and only detects certain scenarios.  The situations searched for with this method are classes of theories where the meson goes free at integer values of $N_c$ and $N_f$.  If possible, one can then attempt to extrapolate to scenarios where the meson goes free at fractional values of $N_f$.  Chiral ring operators with dimension greater than 4/3 cannot appear as ``fundamental" fields in any dual description; they must appear as composite operators built out of an emergent gauge group.  Their R-charges then inform us about the internal structure of the gauge group.

Assume that by varying the size of the gauge group, $N_c$, and the number of fields, $N_f$, one can arrive at a theory with no emergent symmetries, typically a Banks-Zaks fixed point.  At this point, one can trust $a$-maximization of the electric theory.  The standard procedure for detecting emergent $U(1)$ symmetries is that by varying $N_c$ and $N_f$ eventually a chiral ring operator $\OO$ hits the unitarity bound 2/3.  It is then assumed that a single $U(1)$ emerges whose sole result is to make $R_{\OO} = 2/3$.

When a field hits the unitarity bound, there is the possibility that an emergent gauge symmetry has gone IR free.  Let us assume that a field goes free at integer values of $N_f$ and $N_c$.\footnote{In the case of single gauge group dual descriptions, the gauge group can be chosen to go free at integer values of $N_f$ and $N_c$.  In the dual description, the only obstruction to the free description is that the beta function imply IR freedom.  As casmirs are rational numbers, it is always possible to choose integer $N_f$ and $N_c$.  Mixed phases do not have this property as R-charges tend to be irrational in the presence of an additional conformal gauge groups.}  To check for emergent non-Abelian symmetries, one can write down all possible chiral ring operators with R-charge greater than 4/3, $\OO_{R>4/3}$.  Any emergent gauge symmetry must have composite chiral ring operators and it is highly likely that they have large R-charge, especially in the large $N_c$ and $N_f$ limits.  If the gauge symmetry has become IR free, then the R-charge of these large R operators are a integer multiple of 2/3.

The proposed manner by which to detect emergent gauge symmetries is then simple.  When a chiral ring operator hits 2/3 (at integer $N_f$ and $N_c$), check to see if any of the operators  $\OO_{R>4/3}$ have R-charges which are integer multiples of 2/3.  If there are, then the simplest explanation is that an emergent IR free non-Abelian gauge symmetry has appeared.  If only some of the operators have integer multiples of 2/3, then a mixed phase occurs with the free gauge symmetry providing the multiples of 2/3 while the conformal sector provides the rest of the chiral ring operators.

\section{Examples}\label{sec:tests}


\subsection{SQCD}\label{sec:SQCD}

In the electric theory of SQCD we have an $SU(N_c)$ gauge group with $N_f$ fundamental flavors.  For $N_f > 3 N_c$ the Lagrange multiplier for the gauge coupling goes negative, showing that the conformal theory found by $a$-maximization is unphysical.  Instead, the modified $a$-maximization procedure finds the solution of $g = 0$ implying that the theory is IR free.

The meson hits the unitarity bound for $N_f = \frac{3}{2} N_c$.  $N_c$ is chosen even to make this occur at an integer $N_f$ rather than at fractional values.  At this point, the baryonic chiral ring operators have dimension $\frac{N_c}{2} \frac{2}{3}$.  The simplest explanation for the R-charge of the baryons is that there is a magnetic gauge group is going free.  In this case, one can guess that the magnetic gauge group is of size $\frac{N_c}{2}$.  In general, the many different baryonic operators will have many different integers multiplying 2/3 and it will not be obvious what the rank of the gauge group is.

As mentioned in Sec.~\ref{sec:method1}, the magnetic theory has the exact same behavior, except that instead of a chiral operator going free at $N_f = \frac{3}{2} N_c$, there ceases to be a solution.  The lack of a solution combined with the fact that the baryons are integer multiples of $2/3$, indicates that there is likely a non-Abelian gauge symmetry going free.  

We note that the modification to $a$-maximization is symmetric between the magnetic and electric frames of SQCD.  $a$-maximization in its original incarnation was not symmetric under the electric and magnetic descriptions.  It could not see that the magnetic theory needed to be dualized to the electric theory when the meson hit 4/3, but it could see that the electric theory needed to be dualized to the magnetic theory when the meson hit 2/3.

\subsection{SSQCD}

SSQCD is a singlet deformed version of SQCD first considered in \cite{Barnes:2004jj} and further studied in \cite{Amariti:2010sz}.  It contains the feature that mesons can go free before the emergent gauge group goes free.  The theory is 
\begin{center}
\begin{tabular}{c|c|cccc}
&$SU(N_c)$&$SU(N_f)_L$&$SU(N_f)_R$&$SU(N'_f)_L$&$SU(N'_f)_R$  \\
\hline
\hline
&&&\\[-12pt]
$Q$&$\fund$&$\antifund$&$1$&$1$&$1$  \\
$\overline{Q}$&$\antifund$&$1$&$\fund$&$1$&$1$  \\
$Q'$&$\fund$&$1$&$1$&$\antifund$&$1$  \\
$\overline{Q}'$&$\antifund$&$1$&$1$&$1$&$\fund$  \\
$S$&$1$&$1$&$1$&$\fund$&$\antifund$  \\
\end{tabular}
\end{center}
with superpotential
\bea
W = Q' S \overline{Q}'
\eea
A dual description can be obtained by applying Seiberg duality.

Holding $N'_f$ and $N_f$ fixed and varying $N_c$ with $N'_f/N_f$ small enough, nothing becomes free until $N_f + N'_f = \frac{3}{2} N_c$.  At this point, all of the mesons become free and an emergent IR free gauge group appears as described by Seiberg duality.  Much like magnetic SQCD, there is no solution for $N_f + N'_f \le \frac{3}{2} N_c$ indicating that emergent symmetries have appeared.

For $N'_f/N_f$ large enough, one first has the meson $Q \overline{Q}$ going free.  Generically, this crossover occurs for irrational $N_c/N_f$.  e.g. for  $N'_f/N_f = 4$, the cross over occurs at $N_c/N_f = 7-\sqrt{17}$.  At some special values there are rational numbers, for example $N'_f = 15$, $N_f = 3$ and $N_c = 10$.  However, none of the baryons are integer multiples of 2/3, so we expect there not to be any effect beyond a field going free.

After this meson goes free, the a function is corrected by
\bea \label{eq:U1}
a_\mr{new} = a_\mr{old} + N_f^2 (\frac{2}{9} - 3 (2 R_Q - 1)^3 + (2 R_Q - 1))
\eea
At $N_f + N'_f = \frac{3}{2} N_c$, the other two mesons go free.  Before and after accounting for these new mesons going free, there is again no solution.  Combined with an analysis of the baryons, we again determine that there is likely a non-Abelian symmetry going free.

\subsection{A chiral confining Theory}\label{sec:confinement}

To further explore the special nature of the R-charge 4/3 and of baryons with integer multiples of $2/3$, we consider a chiral $s$-confining theory first proposed in \cite{Spiridonov:2009za} using index arguments and later found again with deconfinement \cite{Craig:2011tx}.  The matter content of the theory is 
\begin{center}
\begin{tabular}{c|c|cc}
&$SU(N_c)$&$SU(N_f)$&$Sp(N_f+N_c-4)$  \\
\hline
\hline
&&&\\[-12pt]
$Q$&$\fund$&$\antifund$&$1$  \\
$\overline{Q}$&$\antifund$&$1$&$\fund$  \\
$A$&$\antisym$&$1$&$1$  \\
\end{tabular}
\end{center}
with superpotential
\bea
W = \overline{Q} A \overline{Q}
\eea
Consider $N_f = N_c+ 2$.  The baryon $Q^{N_c}$ has $R=2/3$.  All of the other baryons have R-charge $= n 2/3$, where n is an integer $\ge 2$.  The previous arguments suggest that maybe a gauge group is going free.  Instead of a free magnetic gauge group, this theory $s$-confines into
\begin{center}
\begin{tabular}{c|cc}
&$SU(N_c+2)$&$Sp(2 N_c-2)$  \\
\hline
\hline
&\\[-12pt]
$(Q \overline{Q})$&$\antisym$&$1$  \\
$(Q^{N_c})$&$\fund$&$\antifund$  \\
\end{tabular}
\end{center}
with superpotential
\bea
W = (Q \overline{Q}) (Q \overline{Q}) (Q^{N_c})
\eea
The other baryons $Q^{N_c-2j} A^j$ had $R\ge 4/3$ and are truncated from the chiral ring by non-perturbative effects.

%

\subsection{SQCD with an adjoint} \label{sec:adjoint}

In this subsection, SQCD with an adjoint and a superpotential is briefly discussed.  The values of $N_c$ and $N_f$ where the dual description does not have emergent symmetries are described.

Consider the gauge group $SU(N_c)$ with $N_f$ flavors of quarks and a single adjoint field $X$.  Add the superpotential $W = \text{Tr}{X^{k+1}}$.  Using just $a$-maximization around the theory without a superpotential, we can show that this superpotential is irrelevant until $x = N_c/N_f > \frac{\sqrt{5-8k+5k^2}}{6}$.  Equivalently, this bound can be obtained from the theory with a superpotential and observing when the Lagrange multiplier changes sign.  For asymptotic freedom, we require $x > 1/2$.

The dual for this theory was proposed in \cite{Kutasov:1995ve} and further studied in \cite{Kutasov:1995ss}.  The dual is 
\begin{center}
\begin{tabular}{c|c|cc}
&$SU(k N_f - N_c)$&$SU(N_f)$&$SU(N_f)$  \\
\hline
\hline
&\\[-12pt]
$q$&$\fund$&$\antifund$&$1$  \\
$\overline{q}$&$\antifund$&$1$&$\fund$  \\
$Y$&$adj$&$1$&$1$  \\
$M_j$&$1$&$\fund$&$\antifund$  \\
\end{tabular}
\end{center}
with superpotential
\bea
W = \text{Tr} Y^{k+1} + \sum_{j=1}^k M_j \overline{q} Y^{k-j} q
\eea

We find that in this case, there does not exist any solution with positive Lagrange multipliers until all of the mesons have R-charge less than 4/3.  Before then there is an emergent $U(1)$ symmetry that acts on $M_j$ and $\overline{q} Y^{k-j} q$.  The last meson crosses the $4/3$ boundary for $N_f = \frac{3 N_c}{2 k - 1}$,  thus the dual only becomes valid for $\frac{N_c}{k} \le N_f \le \frac{3 N_c}{2 k +1}$.  Various mesons and $Y$ go free until finally at $N_f = \frac{2 N_c}{2 k - 1}$ the gauge group goes free.  Finally, the dual disappears for $\frac{N_c}{k}$ while the electric theory ceases to have stable vacua.

We point out that much like other methods of detecting emergent symmetries, there are symmetries which escape notice.  In this scenario, the electric theory cannot see when the field $Y$ goes free.  Because $Y$ going free is not related to any unitarity bound or maximum R-charge, the emergence of a $U(1)$ acting on $Y$ cannot be seen using the methods proposed in this paper.

%

\section*{Acknowledgements}

AH thanks B. Wecht for very helpful discussions on the subject and K. Intriligator for helpful communication regarding flow analysis.  AH thanks N. Bao, C. Csaki, S. Kachru, P. Meade, J. Terning, G. Torroba and B. Wecht for comments on the draft and thanks S. Kachru for pestering him for interesting facts so often that AH needed to find one himself.  This work was supported by the US Department of Energy under contract number DE-AC02-76SF00515.  

\appendix

\section{$a$-maximization as a test of duality}\label{sec:appendix}

The extension of $a$-maximization provides a novel check of dualities.  If two theories are dual to each other, ideally one would want to be able to find solutions in one frame when the other frame yields nothing.  However, if one side of the theory never has a solution, then that duality frame is always missing emergent dynamics.  In the worst case, the emergent dynamics could be a quantum modified moduli space invalidating the claim of duality.

Deconfinement is the use of a $s$-confining subsector to give two index tensors proposed in \cite{Berkooz:1995km} and followed up in \cite{Terning:1997jj,Csaki:2004uj,Craig:2011tx}.  It is argued that dualities based on deconfinement can sometimes result in inconsistent CFTs; there are non-trivial dynamics present on the dual side that were previously missed.  Not all results based on deconfinement are inconsistent.  One of the dualities presented in Sec~\ref{sec:tests} can be derived using deconfinement.  

As an example of how this problem can arise, Sec.~\ref{sec:decon} considers the antisymmetric tensor theory of \cite{Csaki:2004uj}.  There it was claimed that for $x > 2.95$, the meson $M$ becomes free.  For $x > 4.09$ the meson $H$ and the entire $Sp$ gauge group goes free.  We will argue that due to negative Lagrange multipliers, no results can be consistently obtained from the proposed dual descriptions.  Thus, one cannot claim the presence of a mixed phase.



\subsection{$s$-confinement and $a$-maximization}
Consider a $s$-confining theory.  For simplicity, consider the gauge group $Sp(N-4)$ with $SU(N)$ flavors of a quark $Y$ where $N$ is even.  After $s$-confinement, the theory has a field $A$ which is an antisymmetric tensor under $SU(N)$ and has superpotential $W = A^{N/2}$.  $a$-maximization on both sides of the duality do not agree with each other.  The emergent dynamics of $s$-confinement prevents $a$-maximization from working on the $Sp$ side.  The proposed extension of $a$-maximization simply does not have a solution in the $Sp$ theory.

Now consider the theory
\begin{center}
\begin{tabular}{c|c|c}
&$Sp(N-3)$&$SU(N)$  \\
\hline
\hline
&\\[-12pt]
$Y$&$\fund$&$\fund$  \\
$Z$&$\fund$&$1$  \\
$P$&$1$&$\antifund$  \\
\end{tabular}
\end{center}
with odd N and superpotential
\bea
W = Y Z P
\eea
After $s$-confinement, the theory is simply an antisymmetric tensor $A = YY$ with no superpotential.  The extra superpotential term removes the dynamical superpotential.  Again, because $s$-confinement was used, one does not expect $a$-maximization to give sensible results on the deconstructed side.  However, there is an interesting coincidence.  The $a$ for the deconstructed theory is 
\bea
a_\mr{deconst} &=& (N-3)(3 (R_Z-1)^3-(R_Z-1)) +  N (3 (R_P-1)^3-(R_P-1))   \nonumber \\
&+& N (N-3) (3 (R_Y-1)^3-(R_Y-1)) \nonumber \\
 &-& g(N-1+N(Y-1)+(Z-1)) - \lambda(2 - R_Y - R_Z - R_P)
\eea
Even though we know the $Sp$ theory is $s$-confining, lets assume that it is conformal and that the superpotential is part of the CFT.  Using the fact that the derivative with respect to $P$,$Z$, $g$ and $\lambda$ vanish, we can solve for those fields.  We find that for $N > 5$, the Lagrange multipliers are always negative.  For the proposed modification of $a$-maximization, this is a problem and there is again no solution due to $s$-confinement.

However, for the original version of $a$-maximization there is no problem and the solution can be plugged back into the a function to get
\bea
a = N(N-1)/2 (3 (2Y-1)^3-(2Y-1)) = a_\mr{confined} (A = 2 Y)
\eea
In other words, an accident has occurred.  Both sides of the duality agree on all R-charges regardless of the fact that $s$-confinement has occurred and $a$-maximization is not expected to give the correct result!

\subsection{deconfinement and mixed phases}\label{sec:decon}

The hope of using deconfinement for mixed phases is that Seiberg duality can successfully prevent $s$-confinement.  $s$-confinement can be hidden from $a$-maximization because considering the $s$-confining theory as conformal still gives the correct results.  $a$-maximization is also asymmetric between the electric and magnetic theories so $s$-confinement can be further hidden from it using Seiberg duality.  The modified $a$-maximization is duality symmetric, so it cannot be fooled by extra dualities.


As an explicit example of deconfinement and mixed phases, consider gauging the $SU(N)$ symmetry adding in matter content to cancel anomalies.  Explicitly, the matter content is 
\begin{center}
\begin{tabular}{c|c|cc}
&$SU(N_c)$&$SU(N_f)$&$SU(N_c+N_f-4)$  \\
\hline
\hline
&&\\[-12pt]
$Q$&$\fund$&$\fund$&$1$  \\
$\overline{Q}$&$\antifund$&$1$&$\fund$  \\
$A$&$\antisym$&$1$&$1$  \\
\end{tabular}
\end{center}
with no superpotential.  This theory was the focus of \cite{Csaki:2004uj}. After deconfining, one can dualize the $SU(N_f)$ gauge group.  This duality masks the $s$-confinement.  Apply duality again, this time to the $Sp$ gauge group.  The resulting theory is
\begin{center}
\begin{tabular}{c|cc|cc}
&$SU(N_f-3)$&$Sp(2 N_f-8)$&$SU(N_f)$&$SU(N_c+N_f-4)$  \\
\hline
\hline
&&\\[-12pt]
$y$&$\antifund$&$\fund$&$1$&$1$  \\
$p$&$\antifund$&$1$&$1$&$1$  \\
$q$&$\fund$&$1$&$\antifund$&$1$  \\
$a$&$\antisym$&$1$&$1$&$1$  \\
$l$&$1$&$\fund$&$1$&$\antifund$  \\
$M$&$1$&$1$&$\fund$&$\fund$  \\
$B$&$1$&$1$&$\fund$&$1$  \\
$H$&$1$&$1$&$1$&$\antisym$  \\
\end{tabular}
\end{center}
with superpotential
\bea
W = M q l y + H l l + B q p + a y y
\eea
Due to the coincidence mentioned before, these two theories have identical values of the $a$ function and R-charges if conformality of gauge groups and marginality of superpotential is imposed.

Now one can forget the procedure used to arrive at the duality and try to compare both sides.  Using the original $a$-maximization procedure, the two sides can be compared and found to agree.   Ignoring Lagrange multipliers but still applying results from unitarity yields the results of \cite{Csaki:2004uj} where they found that for $x >  4.09$ the $Sp$ gauge group went free.

Applying the modification to $a$-maximization, one finds that there is no solution.  Three Lagrange multipliers are negative, $\lambda_{Bqp}$,$\lambda_{ayy}$ and $\lambda_{SU(N_f-3)}$.  Negative Lagrange multipliers are not surprising;  as alluded to in Sec.~\ref{sec:SQCD}, Seiberg duality cannot hide from the extension to $a$-maximization that in one duality frame, the $Sp$ gauge group is $s$-confining.  Because the product gauge theory never has a solution, the deconfined dual always misses emergent dynamics.  

It is not clear in these theories where the emergent dynamics lie.  One would hope that in the large $N_c$ limit that the large number of flavors in the $Sp$ gauge group would make it IR free;  however, depending on the dynamics of the $SU$ sector anything could result.  $s$-confinement would result in no mixed phase; a quantum modified moduli space would result in the duality not matching; higgsing would result in global symmetries or the rank of the $Sp$ gauge group to be wrong; an emergent $U(1)$ would result in unreliable results from $a$-maximization.  Deconfinement does not hide the emergent dynamics.  Unless the final product gauge group has emergent dynamics that are understood, these dual descriptions cannot be used to reliably study the original theory.  In this example, the moduli spaces match on both sides, but a duality cannot be reliably established.  It is in that sense opposite the work of \cite{Brodie:1998vv} where 't hooft anomaly matching did not imply a duality.

%
%
%



\end{document}